\documentclass[epj]{svjour}
\usepackage{latexsym}
\usepackage{graphics}
\newcommand{\be}{\begin{equation}}
\newcommand{\ee}{\end{equation}}
\newcommand{\bea}{\begin{eqnarray}}
\newcommand{\eea}{\end{eqnarray}}
\newcommand{\nn}{\nonumber\\}
\newcommand{\oh}{\frac{1}{2}}

\newcommand{\la}{\langle}
\newcommand{\ra}{\rangle}

\begin{document}
\title{Sum-rules and Goldstone modes from extended RPA theories in Fermi systems
with spontaneously broken symmetries}
\author{D.S. Delion$^{1,2,3}$, P. Schuck$^{4,5}$, and M. Tohyama$^{6}$}
\institute{
$^1$ "Horia Hulubei" National Institute of Physics and Nuclear 
Engineering, \\
407 Atomi\c stilor, POB MG-6, Bucharest-M\u agurele, RO-077125, Rom\^ania \\
$^2$ Academy of Romanian Scientists,
54 Splaiul Independen\c tei, Bucharest, RO-050094, Rom\^ania \\
$^3$ Bioterra University, 81 G\^arlei str., Bucharest, RO-013724, Rom\^ania \\
$^4$ Institut de Physique Nucl\'eaire, F-91406 Orsay CEDEX, France \\
$^5$Laboratoire de Physique et Mod\'elisation des Milieux Condens\'es, 
CNRS and Universit\'e Joseph Fourier, 25 Avenue des Martyrs BP166, 
F-38042 Grenoble C\'edex 9, France \\
$^6$ Kyorin University School of Medicine, Mitaka, Tokyo 181-8611, Japan}

\abstract{
The Self-Consistent RPA (SCRPA) approach is elaborated for cases
with a continuously broken symmetry, this being the main focus of the present article. Correlations beyond standard RPA are summed up  correcting for the quasi-boson approximation in standard RPA. Desirable properties of standard RPA such as fullfillment of energy
weighted sum rule and appearance of Goldstone (zero) modes are kept. 
We show theoretically and, for a model case, numerically that, indeed, SCRPA maintains
all properties of standard RPA for practically all situations of spontaneously broken symmetries.
A simpler approximate form of SCRPA, the so-called renormalised RPA, also has these properties.
The SCRPA equations are first outlined as an eigenvalue problem, but it is also shown how an
equivalent many body Green's function approach can be formulated.}


\maketitle

\section{Introduction}

The present trend in many body physics mostly goes in the direction of a direct numerical treatment of the problem at hand. As examples, one could cite Quantum Monte Carlo (QMC) 
\cite{Nig99,Hol11}
and Density Matrix Renormalisation Group (DMRG) techiques \cite{Pes99,Sch11} which are very successfull.
On the other hand it may also be interesting to make advances in many body theory proper. We will try to  do this in this work employing the equation of motion (EOM) technique (see e.g. \cite{Row68a}). We think that this formalism has not been exploited to its full power in the past. For example,
as is well known, the fullfillment of sum-rules and appearance of Goldstone modes in systems with spontaneously broken symmetries are corner stones of any valuable many body theory. We will dwell on this specific aspect in this work. Unfortunately, going beyond basic approaches like Hartree-Fock (HF)-RPA and/or BCS-QRPA (quasi-particle RPA) \cite{rpae}, it becomes immediately a non-trivial problem to satisfy those properties. There exists the approach of $\Phi$ derivable functionals promoted by Kadanoff and Baym \cite{Kad61}.
However, it is mostly very difficult to implement this numerically beyond HF-RPA (BCS-QRPA), the lowest order approach, because this technique involves in general vertices which depend on more than one energy and, thus, the equations will not only be integral equations in coordinate or momentum spaces but in addition one will have to deal with integrals in energy space what renders most of the problems very complicated. In this work, we want to advocate a different route where we extend the standard HF-RPA (BCS-QRPA) approach including higher correlations without destroying the aforementioned desirable properties. As mentioned, our approach is based on the equation of motion method and leads to a self-consistent version of RPA (SCRPA) which largely overcomes a well known defect of standard RPA, the quasi-boson approximation. There is one point where this defect is easy to trace: RPA, besides excitations, also describes correlations in the ground state. However, fermion occupation numbers used in RPA are the uncorrelated ones, a clear contradiction which is due to the neglect of the Pauli principle. The SCRPA cures this point in using correlated occupation numbers which couple back to the RPA. Additional Pauli corrections as vertex screening are also included. We will show applications to various exactly solvable models and  demonstrate the strong improvement of SCRPA over standard RPA.
We  will treat Fermi systems but we think that the approach can be applied to Bose systems as well.\\

The paper is organized as follows. In Section II we give a short outline of SCRPA. A very simplifying approximation thereof, the renormalised RPA (r-RPA) is presented in Sction III.
In Section IV the Goldstone mode is analyzed and in Section V it is shown that the energy weighted 
sum-rule is fulfilled within SCRPA. In Section VI, we apply SCRPA to the Hubbard and pairing models and we demonstrate numerically with a three level model
that, indeed, the Goldstone (zero) mode appears. In Section VII, we show how SCRPA can equivalently
be formulated with many body Green's functions. In the last Section we draw our conclusions.

\section{Self Consistent Random Phase Approximation (SCRPA)}

As mentioned, our theory will be based on the Equation of Motion (EOM) approach. We will be rather short with the presentation of the formalism, since it has been exposed several times before 
\cite{Row68a,Del05,Duk90,Hir02,Cat96,Sto03}.
For finite systems with discrete levels, the EOM mostly is applied so that an eigenvalue problem results. 
We will follow first this route but later also outline the equivalent Green's function approach which is mostly applied in condensed matter for infinite homogeneous systems. Then for finite systems, if we want to stay within RPA, one may make the following ansatz for an excited state of the system \cite{Row68a}

\begin{equation}
|\nu \rangle = \tilde Q^{\dag}_{\nu}|0\rangle~,
\label{exstate}
\end{equation}

\noindent
where in general $\tilde Q^{\dag}_{\nu}=|\nu\rangle \langle 0|$ is a complicated many body operator, $|0\rangle$ and $|\nu \rangle$ being exact ground and excited states of the system, respectively. To lowest order, we may consider the one body operator

\begin{equation}
\tilde Q^{\dag}_{\nu} = 
\sum_{\alpha \beta}\tilde \chi^{\nu}_{\alpha \beta}a^{\dag}_{\alpha}a_{\beta}~,
\label {rpa-op}
\end{equation}

\noindent
where $a^{\dag}_{\alpha}, a_{\beta}$ are single particle creation and annihilation operators in a general basis but which, to fix the ideas, may be chosen to  correspond to a diagonalisation of the single particle density matrix, the so-called canonical basis. In EOM, one also always supposes that there exists a ground state which is the vacuum to the destructors $\tilde Q_{\nu}$, i.e.,

\begin{equation}
\tilde Q_{\nu}|0\rangle =0~.
\label{killing}
\end{equation}

There are many ways to derive standard HF-RPA. One of the best known is the linearisation of the Time Dependent HF (TDHF) equations, see, e.g., \cite{Rin80}. Here, we want to go a slightly different way. Let us consider the normalised energy weighted sum rule

\begin{eqnarray}
\Omega_{\nu} &=& \frac{1}{2} \frac{\langle 0|[\tilde Q_{\nu},[H,\tilde Q^{\dag}_{\nu}]]|0\rangle}{\langle 0|[\tilde Q_{\nu}, \tilde Q^{\dag}_{\nu}]|0\rangle} 
\nonumber\\&=& 
\frac{\sum_{\mu}(E_{\mu}-E_0)\langle 0|\tilde Q_{\nu}|\mu \rangle |^2}{ \sum_{\mu}\langle 0|\tilde Q_{\nu}|\mu \rangle |^2}~,
\label{sum-rule}
\end{eqnarray}

\noindent
where $E_{\mu}$ and $E_0$ are supposed to be exact eigenvalues of the Hamiltonien $H$ corresponding to the eigenstates $|0\rangle$ and $|\mu\rangle$.
Therefore $\Omega_{\nu}$ can be considered as some average excitation energy and it is this quantity which we want to minimise. As we will see, the summation in (\ref{rpa-op}) must exclude the diagonal terms with indices $\alpha = \beta$. In addition, we want the states $|\nu\rangle $ to be normalised. Accordingly we write for (\ref{rpa-op}) more explicitly

\begin{equation}
Q^{\dag}_{\nu} = \sum_{\alpha > \beta}[X^{\nu}_{\alpha \beta}\delta Q^{\dag}_{\alpha \beta}
-Y^{\nu}_{\alpha \beta}\delta Q_{\alpha \beta}]~,
\label{ex-rpa-op}
\end{equation}

\noindent
where

\begin{equation}
\delta Q^{\dag}_{\alpha \beta} = \frac{A_{\alpha \beta}}{\sqrt{M_{\alpha \beta}}}~,~~A_{\alpha \beta} \equiv a^{\dag}_{\alpha}a_{\beta}
\label{dQ}
\end{equation}

\noindent
are the normalised pair creation operators

\noindent 
with

\begin{equation}
\langle \alpha|\hat M|\beta \rangle \equiv M_{\alpha \beta} = n_{\beta} -n_{\alpha}~,
\label{M}
\end{equation}
 
\noindent
where

\begin{equation}
n_{\alpha} = \langle 0|a^{\dag}_{\alpha} a_{\alpha}|0\rangle~,
\label{occ-numb}
\end{equation}

\noindent
are the single particle occupation numbers. With this choice and with $\langle \nu|\nu \rangle = \langle 0|[Q_{\nu},Q^{\dag}_{\nu}]|0\rangle$, one immediately verifies that with

\begin{equation}
\sum_{\alpha > \beta}\bigg ( |X^{\nu}_{\alpha \beta}|^2 - |Y^{\nu}_{\alpha \beta}|^2 \bigg ) = 1~,
\label{norm}
\end{equation}

\noindent
the excited states $|\nu \rangle$ are normalised under the assumption that we work in the canonical basis where the single particle density matrix only has diagonal elements, that is $\rho_{\alpha \beta} = \langle 0|a^{\dag}_{\beta}a_{\alpha}|0 \rangle = n_{\alpha} \delta_{\alpha \beta}$. From (\ref{M}) and (\ref{dQ}), we see that the configuration $\alpha = \beta $ must be excluded in (\ref{ex-rpa-op}).

With these definitions, we obtain from the minimisation of the sum rule (\ref{sum-rule}) the following eigenvalue equation

\begin{equation}
\left(\matrix{A & B \cr B^* & A^* }\right)
\left(\matrix{X^{\nu}\cr Y^{\nu} }\right)=\Omega_{\nu}
\left(\matrix{ 1& 0 \cr 0 & -1 }\right)
\left(\matrix{X^{\nu}\cr Y^{\nu} }\right)~,
\label{rpa-eq}
\end{equation}

\noindent
where 

\begin{equation}
A_{\alpha \beta \alpha' \beta'} = \langle 0\vert \bigg [ \delta Q_{\alpha \beta},\bigg [ H, \delta Q^{\dag}_{\alpha' \beta'}\bigg ] \bigg ]
\vert 0\rangle~,
\label{A-matrix}
\end{equation}

\noindent
and
\begin{equation}
B_{\alpha \beta \alpha' \beta'} =
- \langle 0\vert \bigg [ \delta Q_{\alpha \beta},\bigg [ H, \delta Q_{\alpha' \beta'}\bigg ] \bigg ]
\vert 0\rangle~.
\label{B-matrix}
\end{equation}

\noindent
For the following it is useful to introduce the quantities

\bea
{\mathcal S} &=& \left(\matrix{A & B  \cr B^* & A^* }\right),~~~
{\mathcal X} = \left(\matrix{X & Y^*\cr Y   & X^* }\right), ~~~
\nn
{\mathcal N}_0 &=& \left(\matrix{1 & 0 \cr 0 & -1 }\right)~.
\label{S-matrix}
\eea

\noindent
We realise that (\ref{rpa-eq}) has exactly the same mathematical structure as standard RPA \cite{Rin80}. For instance, we see that the eigen vectors $\chi^{\nu}$ with components $X^{\nu}$ and $Y^{\nu}$ form a complete orthonormal set.

\noindent
Equation (\ref{rpa-eq}) can then be written in a more  compact form as

\begin{equation}
{\mathcal S}{\mathcal X} = {\mathcal N}_0{\mathcal X}\Omega~,
\label{short-hand-rpa}
\end{equation}

\noindent
where $\Omega$ is a diagonal matrix with real eigenvalues $\Omega_{\nu}, -\Omega_{\nu}$, if ${\mathcal S}$ is positive definite.





The closure relation is 

\begin{equation}
\sum_{\nu}(X^{\nu}_{\alpha \beta}X^{\nu}_{\alpha' \beta'}-Y^{\nu}_{\alpha \beta}Y^{\nu}_{\alpha' \beta'}) = \delta_{\alpha \alpha'}\delta_{\beta \beta'}~.
\label{closure-bis}
\end{equation}

\noindent
The orthonormality relations allow us to invert the operator (\ref{ex-rpa-op}). For $\alpha > \beta$ we have

\bea
a^{\dag}_{\alpha}a_{\beta} &=& \sqrt{M_{\alpha \beta}}
\sum_{\nu}(X^{\nu *}_{\alpha \beta}Q^{\dag}_{\nu} + Y^{\nu *}_{\alpha \beta}Q_{\nu})
\nn
a^{\dag}_{\beta}a_{\alpha} &=& \sqrt{M_{\alpha \beta}}
\sum_{\nu}(X^{\nu }_{\alpha \beta}Q_{\nu} + Y^{\nu }_{\alpha \beta}Q^{\dag}_{\nu})~.
\label{inversion}
\eea

\noindent
With (\ref{killing}), it then follows that the density matrix $\langle 0|a^{\dag}_{\alpha}a_{\beta}|0 \rangle$ only has diagonal elements, as it was introduced already after (\ref{norm}).\\

It can immediately be verified that, if all expectation values in (\ref{rpa-eq}) are evaluated with the HF ground state, then the standard RPA equations \cite{Rin80} are recovered with, in particular, only $X^{\nu}_{ph}$ and $Y^{\nu}_{ph}$ amplitudes surviving where the indices $p(h)$ stand for 'particle (hole)', i.e., indices above (below) the Fermi energy. The equations (\ref{rpa-eq}) are, however, much more general and it is obvious that, if the expectation values in (\ref{rpa-eq}) are evaluated with the RPA ground state obeying (\ref{killing}), then the matrices $A$ and $B$ will depend in a complicated nonlinear way on the amplitudes $X$ and $Y$. This we will call the Self-Consistent RPA (SCRPA).

Before we come to the explicit evaluation of the matrix elements $A, B$ in (\ref{rpa-eq}) and to the discussion of spontaneously broken symmetries together with sum-rules, Goldstone modes, etc., we first shall deal with the so far open but very important question of the optimal single particle basis. As usual, we will obtain this from the minimisation of the ground state energy with respect to the basis. We will show that this minimisation is equivalent to the following additional equation of motion

\begin{equation}
\langle 0|[H, Q^{\dag}_{\nu}]|0\rangle = \langle 0|[H, Q_{\nu}]|0\rangle =0~,
\label{rpa-mf}
\end{equation}

\noindent
which obviously is correct if $|0\rangle$ is an eigenstate of $H$. Because there are as many operators $Q^{\dag}_{\nu}, Q_{\nu}$ as there are components $a^{\dag}_{\alpha}a_{\beta}, a^{\dag}_{\beta}a_{\alpha}$, we also can write for (\ref{rpa-mf})

\begin{equation}
\langle 0|[H,a^{\dag}_{\alpha}a_{\beta} ]|0\rangle = \langle 0|[H, a^{\dag}_{\beta}a_{\alpha}  ]|0\rangle =0~,
\label{rpa-mfbis}
\end{equation}

\noindent
where we again recall our convention $\alpha > \beta$. Equations (\ref{rpa-mfbis}) are of the one body type and one can directly verify that with a Slater determinant as ground state, they reduce to the HF equations. 

Minimising the ground state energy with respect to the basis, let us first write the hamiltonian in our general single particle basis with greek indices more explicitly. Supposing that the hamiltonian is originally given in the basis of plane waves which are written in the sought for basis as $c^{\dag}_k = \sum_{\alpha}R^*_{\alpha, k}a^{\dag}_{\alpha}$
where $k$ includes momenta, spin, and isospin, the hamiltonian with two body interactions in the new basis reads

\bea
H&=& \sum_{k \alpha \beta}e_kR^*_{\alpha, k}R_{\beta, k}a^{\dag}_{\alpha}a_{\beta}
\nn&+&
\frac{1}{4}
\sum \bar v_{k_1k_2k_3k_4}
R^*_{\alpha, k_1}R^*_{\beta, k_2}R_{\gamma, k_3}R_{\delta, k_4}
a^{\dag}_{\alpha}a^{\dag}_{\beta}a_{\delta}a_{\gamma}~,
\nn
\label{H}
\eea

\noindent
with $e_k = k^2/(2m)$ the kinetic energy and $\bar v_{k_1k_2k_3k_4}$ the antisymmetrised matrix element of the two body interaction. The  corresponding variational equation are obtained from

\begin{equation}
\frac{\partial}{\partial R^*_{\alpha, k}}
\bigg (\langle 0|H|0\rangle -\sum_{\beta} E_{\beta}
\sum_kR^*_{\beta, k}R_{\beta, k} \bigg ) = 0~,
\label{var-eq1}
\end{equation}

\begin{equation}
\frac{\partial}{\partial R_{\alpha, k}}\bigg (\langle 0|H|0\rangle -\sum_{\beta} E_{\beta}
\sum_kR^*_{\beta, k}R_{\beta, k} \bigg ) = 0~,
\label{var-eq2}
\end{equation}

\noindent
where, as usual, we ensured with Lagrange multipliers that the transformation is unitary. 
For the common situation where the transformation is real, this yields the following eigenvalue problem

\begin{equation}
\sum_{k'}{\mathcal H}_{kk'}R_{\alpha, k'} = E_{\alpha}R_{\alpha, k}~,
\label{MF}
\end{equation}

\noindent
with

\begin{equation}
{\mathcal H}_{k_1k_2}= h^{MF}_{k_1k_2}+ \frac{1}{2}\sum_{k_3k_4k_5}\bar v_{k_1k_3k_4k_5}C_{k_4k_5k_2k_3}~,
\label{MF-H}
\end{equation}

\noindent
and the mean field (MF) hamiltonian given by

\begin{equation}
h^{MF}_{kk'} = e_k\delta_{kk'}+ \sum_{k_1k_2}\bar v_{kk_1k'k_2}\rho_{k_2k_1}~.
\label{hMF}
\end{equation}

\noindent
The single and two particle density matrices corresponding to the operators $\hat \rho$ and $\hat C$, respectively, are 

\bea
\rho_{k_1k_2} &=& \langle 0|a^{\dag}_{k_2}a_{k_1}|0\rangle
\nn
C_{k_1k_2k_3k_4} &=& \langle 0|a^{\dag}_{k_1}a^{\dag}_{k_2}a_{k_4}a_{k_3}|0\rangle
-(\rho_{k_1k_3}\rho_{k_2k_4}
\nn
&-& \rho_{k_1k_4}\rho_{k_2k_3})~,
\label{DM1,2}
\eea

\noindent
It is now easy to show that (\ref{rpa-mfbis})    and (\ref{MF}) are equivalent.\\

Supposing that we work in this optimised single particle basis with greek indices, we obtain for the SCRPA matrix in (\ref{short-hand-rpa}) \cite{Duk98}

\begin{eqnarray}
\tilde {\mathcal S}_{\alpha \alpha', \beta \beta'} &=& \sqrt{M_{\alpha \alpha'}}(e_{\alpha \beta}\delta_{\alpha' \beta'} - e_{\alpha' \beta'}\delta_{\alpha \beta})\sqrt{M_{\beta \beta'}}
\nonumber\\ 
&+& \delta_{\alpha' \beta'}\frac{1}{2}\sum_{\gamma \gamma' \gamma''} \bar v_{\alpha \gamma \gamma' \gamma''}C_{\gamma' \gamma'' \gamma \beta}\nonumber\\
&-&\delta_{\alpha \beta}\frac{1}{2}\sum_{\gamma \gamma' \gamma''}\bar v_{\gamma \gamma' \alpha' \gamma''}C_{\beta' \gamma'' \gamma \gamma'}\nonumber\\
&+& M_{\alpha \alpha'}\bar v_{\alpha \beta'\alpha'\beta}M_{\beta \beta'}\nonumber\\ 
&+& \sum_{\gamma \gamma'}(\bar v_{\alpha \gamma \beta \gamma'}C_{\beta' \gamma' \alpha' \gamma} + \bar v_{\beta' \gamma \alpha' \gamma'}C_{\alpha \gamma' \beta \gamma})\nonumber\\
&-&\frac{1}{2}\sum_{\gamma \gamma'}(\bar v_{\alpha \beta' \gamma \gamma'} C_{\gamma \gamma' \alpha' \beta} + \bar v_{\gamma \gamma' \alpha' \beta}C_{\alpha \beta' \gamma \gamma'}~),
\nn
\label{s-matrix}
\end{eqnarray}

\noindent
where we introduced the non-diagonal kinetic energies $e_{\alpha \beta}$ and $\tilde {\mathcal S} = \hat M^{1/2}{\mathcal S}\hat M^{1/2}$ and supposed that the single particle density matrix is also diagonal, as stated after (\ref{inversion}). It seems, however, clear that density matrix and single particle Hamiltonian (\ref{hMF}) cannot be diagonal
simultaneously. This contradiction stems from the fact that the killing condition (\ref{killing}) cannot be satisfied exactly. With (\ref{ex-rpa-op}) it is only satisfied to good approximation choosing, e.g., the ground state wave function to be the one of Coupled Cluster Theory (CCT) at SUB2 level, as this is discussed in detail in ref.
\cite{Jem13}.

In order to establish selfconsistency, we must express the matrices of $\hat M$ and $\hat C$ by the amplitudes $X, Y$. To this end, we write

\begin{equation}
\delta_{\alpha \beta'}\delta_{\beta \alpha'}n_{\beta}(1-n_{\alpha}) + C_{\alpha \beta \alpha' \beta'} = \langle a^{\dag}_{\alpha'}a_{\alpha}a^{\dag}_{\beta'}a_{\beta}\rangle~,
\label{C-ph}
\end{equation}

\noindent
with abreviation $\langle ... \rangle = \langle 0|...|0\rangle$.
With (\ref{inversion}) and the killing condition (\ref{killing}), we can express this correlation function for $\alpha \ne \alpha'$ and $\beta \ne \beta' $ by $X$ and $Y$ together with single particle occupation numbers $n_{\alpha}$. 


Concerning the s.p. occupation numbers, we refer to expressions derived earlier in the literature employing the so-called number operator method, see e.g., \cite{Cat96}. We also have  derived these expressions from the Coupled Cluster (CC) wave function at level SUB2 \cite{Jem13}.
In model cases it has been shown that this wave function fullfills the killing relation 
(\ref{killing}) to very good accuracy \cite{Hir02} as was mentioned already above. The equations read

\begin{equation}
n_h = 1 -\frac{1}{2}\sum_{p, \nu}M_{ph}|Y_{ph}^{\nu}|^2 ~~ = 1-\frac{1}{2}\sum_p \langle a^+_pa_ha^+_ha_p\rangle~,
\label{n_h}
\end{equation}

\begin{equation}
n_p = \frac{1}{2}\sum_{h, \nu}M_{ph}|Y_{ph}^{\nu}|^2~~ = \frac{1}{2} \sum_h \langle a^+_pa_ha^+_ha_p\rangle~,
\label{n_p}
\end{equation}

\noindent
or

\begin{equation}
M_{ph} = 1 -\frac{1}{2}\sum_{p, \nu}M_{ph}|Y_{ph}^{\nu}|^2 - \frac{1}{2}\sum_{h, \nu}M_{ph}|Y_{ph}^{\nu}|^2~.
\label{Neq}
\end{equation}

\noindent
Let us remind that $p(h)$ stand for  greek indices above (below) the Fermi surface. \\

The correlation functions $\langle a^{\dag}_{\alpha}a_{\alpha}a^{\dag}_{\beta}a_{\beta} \rangle$ need extra care. If $\alpha = \beta$, there is no problem since the expression reduces to 
a single particle occupation number which we just treated above. In the case $\alpha \ne \beta$, we can write (please be aware that all greek quantum numbers represent just a single quantum state)

\begin{equation}
\langle a^{\dag}_{\alpha}a_{\alpha}a^{\dag}_{\beta}a_{\beta} \rangle = n_{\alpha} - \langle a^{\dag}_{\alpha}a_{\beta}a^{\dag}_{\beta}a_{\alpha}\rangle, ~~~~\alpha \ne \beta~.
\label{nn}
\end{equation}

\noindent
Again, we can now express the two body correlation function in (\ref{nn}) by the $X, Y$ amplitudes and all single and two body density matrices can be fully included in a selfconsistent solution of the SCRPA equations (\ref{rpa-eq}).

The efficiency of the method has been demonstrated in several earlier publications 
\cite{Del05,Duk90,Hir02,Cat96,Sto03}. Examples will be given below including one with a broken symmetry with the appearence of a Goldstone (zero) mode which is the main subject of this paper. 

\section{Renormalised RPA}

A very much simplified version of SCRPA consists in the so-called renormalised RPA (r-RPA). It simply is obtained in discarding in (\ref{s-matrix}) and (\ref{C-ph}) the two-body correlations $\hat C$, keeping, however, the correlated occupation numbers (\ref{n_h}-\ref{Neq}).

\begin{equation}
{\mathcal H}^{(r)}_{k_1k_2}= h^{\mbox{MF}}_{k_1k_2}
\label{H^r}
\end{equation}

\begin{eqnarray}
\tilde {\mathcal S}^{(r)}_{\alpha \alpha', \beta \beta'} &=& (E_{\alpha}-E_{\alpha'})M_{\alpha \alpha'}\delta_{\alpha \beta}
\delta_{\alpha'\beta'}\nonumber\\ 
&+& M_{\alpha \alpha'}\bar v_{\alpha \beta'\alpha'\beta}M_{\beta \beta'}
\nn
\label{s-matrix1}
\end{eqnarray}

Since the correlated occupation numbers are rounded near the Fermi surface, we again can consider an RPA operator with all configurations besides diagonal generators. This feature preserves the property of fullfillment of sum rule and appearance of Goldstone (zero) modes to be discussed below. In a way, r-RPA is the most direct extension of standard RPA: we know that standard RPA accounts for correlations. However, as already mentioned, the RPA equations are set up, e.g., with uncorrelated occupation numbers. In r-RPA this contradiction is lifted. The performance of r-RPA is somewhat between standard and SCRPA \cite{Hir02,Cat96,Del05}

\section{Goldstone modes}

It is well established that the standard HF-RPA (BCS-QRPA) approach exhibits a Goldstone (zero) mode if the HF solution corresponds to a continuously broken symmetry. For finite systems, with discrete quantum numbers, one mostly talks about a zero or spurious mode. For example, for nuclei and other selfbound Fermi systems like 3He droplets, HF always breaks translational invariance and the corresponding RPA shows a zero mode \cite{Gol61,Bla86,Rin80} which corresponds to a coherent displacement of the whole system. In trapped cold atom gases the corresponding mode is the Kohn mode where the atom cloud oscillates coherently in an external harmonic trapping potential \cite{Koh61,Rei01}.
Within BCS-QRPA, one obtains in infinite matter because of broken particle number $U(1)$ symmetry a Goldstone mode, the so-called Bogoliubov-Anderson mode \cite{And58,Com06,Urb14}
Also in finite superfluid systems, like many superfluid nuclei, zero modes appear \cite{Bes66,Kha04}.

As mentioned, these Goldstone (zero) modes reflect basic principles of quantum mechanics and it is very important not to destroy this property in theories which go beyond the HF-RPA scheme. In this respect, what is the situation with the SCRPA approach outlined above? As we see from (\ref{sum-rule}), the crucial point is that the $Q^{\dag}$ operator can represent the symmetry operator (let us call it $\hat P$) in question as a particular solution of SCRPA equations and that the relation $[H, \hat P]=0$ is not destroyed in the course of applying the formalism. In standard RPA, only the $ph$ and $hp$ components of the symmetry operator enter the equations. One can ask the question why the zero mode appears nonetheless. The answer comes from the fact that even if one included the $pp$ and $hh$ matrix elements of $\hat P$ into the RPA, these elements become, in the HF basis, completely decoupled from the rest of the RPA equations. Therefore, in standard RPA, it is as if the totality of the symmetry operator were included  and, thus, the Goldstone mode is present nonetheless. In SCRPA {\it all} components of the symmetry operator, besides the diagonal ones, are present and they all play a role. Therefore, one may think that if $\hat P_{\alpha \alpha}=0$, then in any case the Goldstone mode will come in the solution. In this respect, we must remember that SCRPA joins HF-RPA in the weak coupling limit. Therefore, the inclusion of the generalised mean field (MF) equation (\ref{MF}) must also be assured for the appearance of a zero mode. In analogy to the HF-RPA scheme, one may talk of the MF-SCRPA scheme to imply a zero mode in the broken symmetry case. In the next section, we will demonstrate explicitly the appearance of the Goldstone mode with an application to a model case.\\

All this holds true under the condition that $\hat P_{\alpha \alpha} =0$. Many symmetry
operators have this property. This is the case for the linear momentum because of its odd parity. So the zero mode corresponding to translational motion will come in selfbound systems like nuclei or 3He droplets. Equally the above mentioned Kohn mode in trapped cold atom systems. Some systems may be deformed and then rotational symmetry is broken. The angular momentum operator has no diagonal elements either because it is not time-reversal invariant.

More subtle is the question of pairing which is one of the broken symmetries often encountered in Fermi systems. Indeed, in this case, the symmetry operator is the particle number operator $\hat N = \sum_{\alpha}a^{\dag}_{\alpha}a_{\alpha}$. In quasiparticle representation $a^{\dag}_{\alpha} = u_{\alpha}q^{\dag}_{\alpha} -v_{\alpha}q_{-\alpha}$ this becomes

\begin{eqnarray}
\hat N &=& \sum_{\alpha}[v_{\alpha}^2 + u_{\alpha}^2q^{\dag}_{\alpha}q_{\alpha} - v_{\alpha}^2q^{\dag}_{-\alpha}q_{-\alpha}
\nonumber\\
&-& u_{\alpha}v_{\alpha}(q^{\dag}_{\alpha}q^{\dag}_{-\alpha}+ q_{-\alpha}q_{\alpha})].
\label{Nop}
\end{eqnarray}

\noindent
which contains a hermitian diagonal piece which cannot be included into the (quasi-particle) RPA operator $Q^{\dag}$ as we discussed already above. In this case the standard BCS-QRPA scheme \cite{Rin80}, analogous to HF-RPA in the non-superfluid case, shows, as well known \cite{Bes66,Kha04}, a zero mode because the so-called scattering terms $(q^{\dag}q)$ drop out of the QRPA equations. However, with the Self-Consistent QRPA(SCQRPA) this is not the case and in general the zero mode will be absent. As it turns out, this is only a problem for finite systems. For homogeneous infinite matter we have as QRPA operator

\begin{eqnarray}
Q^{\dag}_{{\bf q}} &=&\sum_{{\bf k}>0}[X^{{\bf q}}_{{\bf k}}q^{\dag}_{{\bf q} -{\bf k}}q^{\dag}_{{\bf k}} -Y^{{\bf q}}_{{\bf k}}q_{{\bf q} -{\bf k}}q_{{\bf k}}\nonumber\\
&+&Z^{(+){\bf q}}q^{\dag}_{{\bf q}+{\bf k}}q_{{\bf k}}-Z^{(-){\bf q}}_{{\bf k}}q^{\dag}_{{-\bf k}}q_{{\bf q}-{\bf k}}]
\label{QRPAop}
\end{eqnarray}

\noindent
where ${\bf q}$ is the c.o.m. momentum of the pair excitation (we suppressed spin indices). So, as long as ${\bf q}$ is not strictly zero, we can include the scattering terms and one will approach the zero mode in the limit as 
${\bf q} \rightarrow 0$. One realises that in finite systems with discrete levels, the zero mode is absent in MF-SCQRPA and one will have to extend the theory. This can be done in including to the QRPA operator, besides the one body sector, the two body sector what will allow the inclusion of diagonal hermitian one body pieces in the QRPA operator and the zero mode is saved. Some works on including the two-body sector can be found, e.g., in \cite{Tohy15} and references in there. However, it shall not be the task of this work to enter into the more complicated structure of this approach. It may be investigated in future work. For the moment, let us be satisfied that the Bogoliubov-Anderson mode \cite{And58} appears in the infinite matter case.

Another particular case occurs, if we write the RPA operator with collective $ph$ operators as this may be intuitively the case when s.p. states are degenerate and which often reduces the dimension of the SCRPA equations drastically. The price to be paid are some extra complications with certain two body correlation functions as we will discuss with the application to a model just below. However, in general, the so-called m-scheme where one considers quantum state by quantum state is preferable because the complications with the evaluation of the density matrices are absent. For example with the ansatz (\ref{QRPAop}) the latter is the case.

\section{Sum-rules}

We show that the energy weighted sum rule, given by the following identity

\begin{equation}
\sum_{\nu}(E_{\nu}-E_0)|\langle \nu|F|0\rangle|^2 = \frac{1}{2}\langle 0|[F,[H,F]]|0\rangle
\label{srul0}
\end{equation}

\noindent
is fullfilled within SCRPA. Here $|\nu\rangle$ is a complete set of eigenstates and $F$ is a one body operator

\begin{equation}
F=\sum_{\alpha \beta} f_{\alpha \beta}A_{\alpha \beta}
\label{F-op}
\end{equation}

\noindent
where $A_{\alpha \beta}$ is defined in (\ref{dQ}). One can show that the identity (\ref{srul0})
 is automatically fullfilled
if one considers that $|\nu\ra$ is the set of SCRPA or r-RPA eigenstates.
By using the inverse transformation of the basis operators
$A_{\alpha \beta}$  one obtains
\bea
\label{srul1}
&&\sum_{\nu}(E_{\nu}-E_{0})|\la\nu|F|0\ra|^2
\nn&=&
\sum_{\nu}(E_{\nu}-E_{0})|\la0|[Q_{\nu},F]|0\ra|^2
\nn
&=&\sum_{\nu}(E_{\nu}-E_{0})
|\sum_{\alpha\beta}f_{\alpha\beta}M_{\alpha\beta}^{1/2}
(X_{\alpha\beta}^{\nu}+Y_{\alpha\beta}^{\nu})|^2~.
\nn
\eea
Using the general system of RPA equations with excitation energy
$\Omega_{\nu}=E_{\nu}-E_0$ one gets the following identity
\bea
\label{srul2}
&&\sum_{\nu}(E_{\nu}-E_{0})|\la\nu|F|0\ra|^2
\nn&=&
\sum_{\alpha\beta}f_{\alpha\beta}M_{\alpha\beta}^{1/2}
\sum_{\gamma\delta}f_{\gamma\delta}M_{\gamma\delta}^{1/2}
(A_{\alpha\beta,\gamma\delta}
-B_{\alpha\beta,\gamma\delta})~.
\nn
\eea
On the other hand by splitting the summations into $\alpha > \beta$
and $\alpha < \beta$ parts and using the definitions of the RPA matrices
one obtains that the right-hand side of the Eq. (\ref{srul2})
has the same form
\bea
\label{srul3}
&&{\oh}\la0|[F,[H,F]]|0\ra
\nn&=&
\sum_{\alpha\beta}\sum_{\gamma\delta}f_{\alpha\beta}f_{\gamma\delta}
{\oh} \la0|[A_{\alpha\beta},[H,A_{\gamma\delta}]]|0\ra~.
\eea

We would like to emphasise that this sum rule also remains fulfilled
 in the case of
broken  symmetry, in spite of the fact that for a   zero energy eigenstate the amplitudes $X,Y$ diverge.
\section{Applications to models}

\subsection{The Hubbard model}

We shortly want to outline how our formalism works in the Hubbard 
model and present some results. In momentum space, the Hamiltonian is given by
(see, e.g., \cite{Mohsen})
\begin{eqnarray}
\label{Ham-HubbMod}
H=&&\sum_{{\bf k},\sigma} (\epsilon_k - \mu)\hat n_{{\bf k}, \sigma}
\\
&&+\frac{U}{2N}
\sum_{{\bf k},{\bf p},{\bf q}, \sigma} a^{\dag}_{{\bf k},\sigma}a_{{\bf k}+{\bf q},
\sigma} a^{\dag}_{{\bf p}, -\sigma}a_{{\bf p}-{\bf q}, -\sigma}
\nonumber
\end{eqnarray}

\noindent
where $\hat n_{{\bf k},\sigma}=a^{\dag}_{{\bf k},\sigma}a_{{\bf k},\sigma}$ is the 
occupation number operator and the single particle energies are given 
by $\epsilon_{{\bf k}} = -2t\sum_{d=1}^D cos(k_d)$ with the lattice spacing 
set to unity. It is convenient to transform the fermion creation and annihilation 
operators $a^{\dag},a$ to HF quasi-particle operators. In 1D, we have

\begin{equation}
a_{h,\sigma} = b^{\dag}_{h, \sigma}~,~~~~a_{p,\sigma} = b_{p,\sigma}
\label{transf-atobHubb}
\end{equation}

\noindent
where $h$ and $p$ are momenta below and above the Fermi momentum, 
respectively, so that $b_{k,\sigma}|\mbox{HF}\rangle =0$ for all $k$ where 
$|\mbox{HF}\rangle$ is the Hartree-Fock ground state in the plane wave basis.
Introducing the operators

\begin{equation}
\tilde n_{k,\sigma} = b^{\dag}_{k,\sigma}b_{k,\sigma} ~,
\label{occpNumHubb}
\end{equation}

\begin{equation}
J^-_{ph,\sigma}=b_{h,\sigma}b_{p,\sigma}~,~~~~J^+_{ph,\sigma}=(J^-_{ph,\sigma})^{\dag}
\label{quasiPartOp}
\end{equation}


\noindent
we write for the RPA operator



\begin{equation}
Q^+_{q,\nu}=\sum_{ph,\sigma}[  \bar X^{\nu}_{ph,\sigma}J^+_{ph,\sigma} - \bar Y^{\nu}_{ph, -\sigma}
J^-_{ph, -\sigma}] 
\label{op-rpa}
\end{equation}





With our Eqs. (\ref{n_h}, \ref{n_p}) for the evaluation of the occupation numbers, we also verify 
straightforwardly that

\begin{eqnarray}
\langle n_{p\sigma}\rangle = \langle \tilde{n}_{p\sigma}\rangle &=& \sum_{h}\langle J^{+}_{ph\sigma} J^{-}_{ph\sigma}\rangle
\nonumber \\
& = & \sum_{h,\nu} (1-\langle M_{ph\sigma}\rangle) |Y^\nu_{ph\sigma}|^2, 
\nonumber \\
\langle n_{h\sigma}\rangle =1&-& \langle \tilde{n}_{h\sigma}\rangle =  1- \sum_{p}\langle J^{+}_{ph\sigma} J^{-}_{ph\sigma}\rangle
\nonumber \\
& = & 1- \sum_{p,\nu} (1-\langle M_{ph\sigma}\rangle) |Y^\nu_{ph\sigma}|^2
\label{expOccupNumb-Hubb}
\end{eqnarray}
\noindent
where $\langle M_{ph\sigma}\rangle = \langle \tilde{n}_{h\sigma}\rangle + \langle \tilde{n}_{p\sigma}\rangle$, 
and  more complicated expressions for the quadratic terms.
Those expressions are  the same as derived in our earlier publication \cite{Mohsen}.\\


\noindent
{\it The Hubbard molecule.} In the Hubbard model, the results are again quite promising. For example the half filled 2-sites problem, the so-called Hubbard molecule, is solved exactly. This is not a totally trivial result. For example the well known GW approximation fails in this respect \cite{Mohsen}.\\

\begin{figure}
\resizebox{0.50\textwidth}{!}{%
\includegraphics{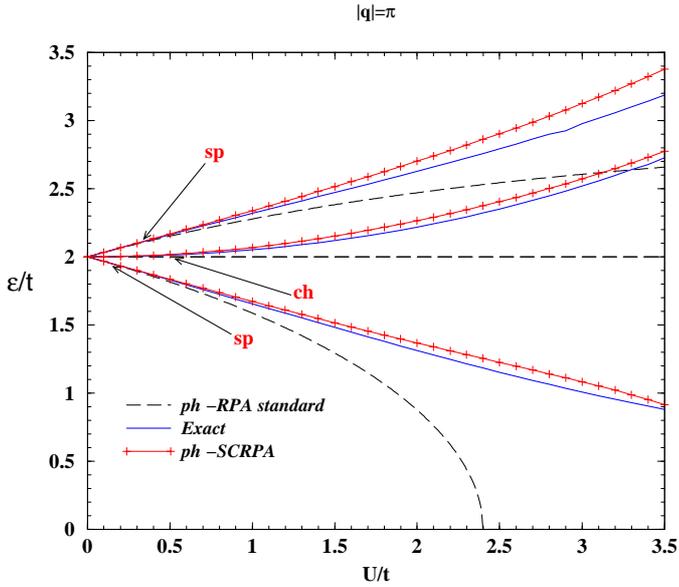}}
\caption{Excitation spectrum for the transfer $|q| = \pi$ as a function of $U/t$ 
of the 6-sites half-filled Hubbard chain. 'sp' and 'ch' stand for spin and charge response, respectively.}
\label{Mohsen4} 
\end{figure}

\begin{figure}
\resizebox{0.50\textwidth}{!}{%
\includegraphics{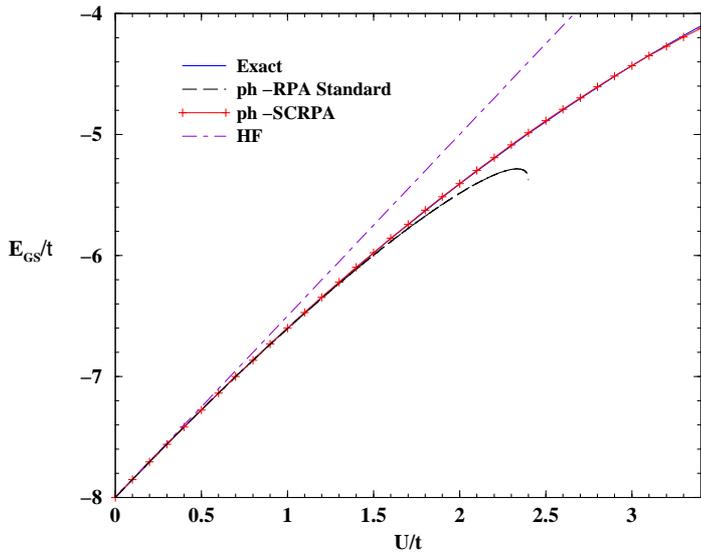}}
\caption{Ground state energy as a function of $U/t$ for the 6-sites half-filled
 Hubbard chain with SCRPA, standard RPA, HF,  and exact solution.}
\label{Mohsen7} 
\end{figure}

\noindent 
{\it The half-filled 6-sites chain.} We show in Fig.\ref{Mohsen4}, for a choice, the excitation spectrum for the momentum transfer $|q|=\pi$. The abreviations 'ch' and 'sp' stand for 'charge', i.e. spin $S=0$ and for 'spin', that is $S,M =1,0$ excitations, respectively. The results for $|q| = 2\pi/3, \pi /3$ are of similar quality \cite{Mohsen}. In Fig.\ref{Mohsen7}, we show the ground state energy. There is good agreement with the exact solution and it presents a maximum error of about $0.8$ percent at $U/t = 3.5$. In that figure, we also display the results of HF and standard RPA. We can appreciate the gain in precision with SCRPA.

\begin{figure}
\resizebox{0.50\textwidth}{!}{%
\includegraphics{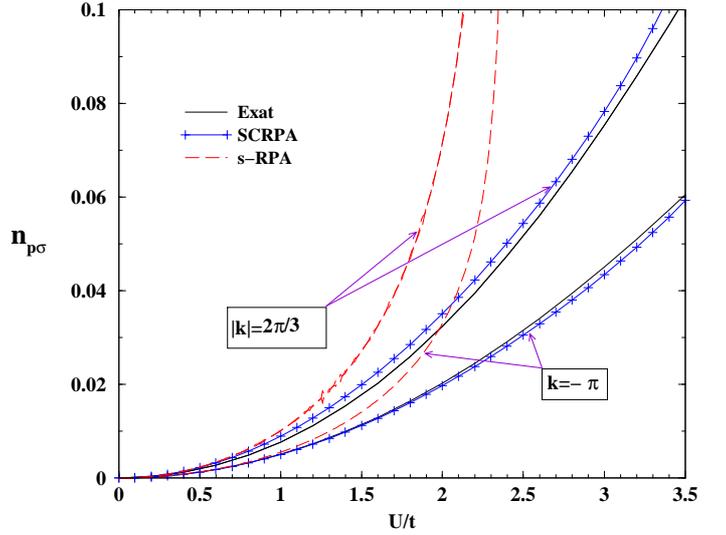}}
\caption{Particle occupation numbers for momenta $k=\pi/3$ and ${\bf k}=-\pi$ in SCRPA
 and standard RPA (s-RPA) for the 6-sites half filled Hubbard chain as a function of $U/t$.}
\label{Mohsen10} 
\end{figure}

\begin{figure}
\resizebox{0.50\textwidth}{!}{%
\includegraphics{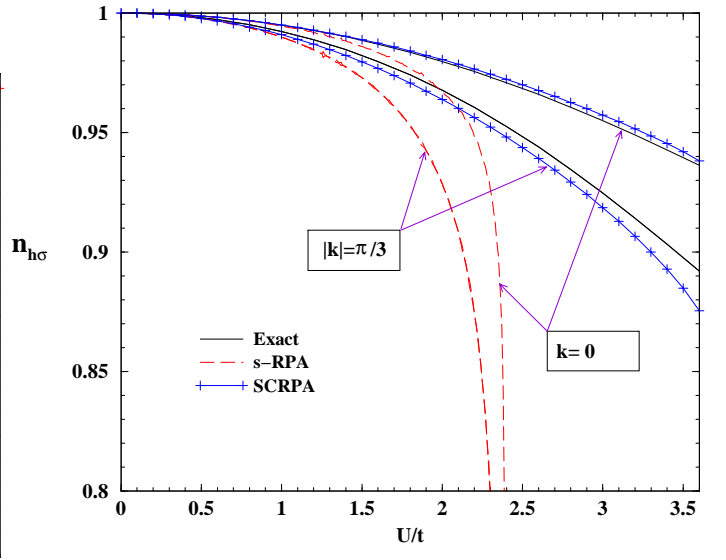}}
\caption{Same as Fig. \ref{Mohsen10} but for hole occupations.}
\label{Mohsen11} 
\end{figure}

In Figs. \ref{Mohsen10} and \ref{Mohsen11} , we show the occupation numbers, see Eq.(\ref{expOccupNumb-Hubb}). We see that with SCRPA they compare very well with the exact values and are very much improved over the corresponding values from standard RPA (s-RPA). It is worth mentioning that particle number is conserved, i.e., what is depleted below the Fermi surface is exactly replaced by non-zero values above the Fermi surface. In the macroscopic limit, this implies that the Luttinger theorem (see, e.g., \cite{Luttinger}) is respected. Let us also mention that the 1D Hubbard model has been solved with renormalised RPA in the infinite system limit with interesting results particularly in the strong coupling limit, \cite{Schaefer}.

\subsection{Pairing model}

The pairing or picket fence model (PFM) is the only one where in  the past the 
SCRPA scheme could be applied for the first time without any approximation and without the 
explicit knowledge of the vacuum \cite{Hir02}. This stemmed from the fact that in this 
particular model with the Hamiltonian

\begin{equation} 
H= \sum_i \varepsilon_i N_i  + V\sum_{ik} P^+_i P_k
\label{Ham-PFM}
\end{equation}

\noindent
and N two fold degenerate  equidistant levels labeled with the index 'i', the 
occupation number operators can exactly be expressed by the product of two 
fermion pair operators, that is

\begin{equation}
N_i=2P_i^{\dag}P_i
\label{Numb-PFM}
\end{equation}

\noindent
with $N_i = a_{i+}^{\dag}a_{i+} + a_{i-}^{\dag}a_{i-}$ and $P_i^{\dag} = 
a_{i+}^{\dag}a_{i-}^{\dag}$. It is seen that the pair operators are the ones which 
enter the Bogoliubov transformation of fermion pairs in the pp-SCRPA 

\begin{equation}
Q^+_{\alpha} = \sum_p \bar X_p^{\alpha}P^+_p + \sum_h \bar Y_h^{\alpha} P^+_h 
\label{Qalpha-PFM}
\end{equation}

\noindent
and, therefore, with (\ref{killing}) 
it was possible in \cite{Hir02} to 
calculate $\langle N_i\rangle$ and $\langle N_iN_j\rangle$ completely 
selfconsistently and without the use of any procedures external to the SCRPA 
ones. We also remark that the evaluation of $\langle N_iN_j\rangle$ 
necessitates the knowledge of four particle correlation functions what makes 
the approach rather heavy. However, factorisation $\langle N_iN_j\rangle 
\sim \langle N_i\rangle \langle N_j\rangle$ turned out to work quite well, 
thus strongly simplifying the expressions \cite{Duke-PLB}.\\

\begin{figure}
\resizebox{0.50\textwidth}{!}{%
\includegraphics{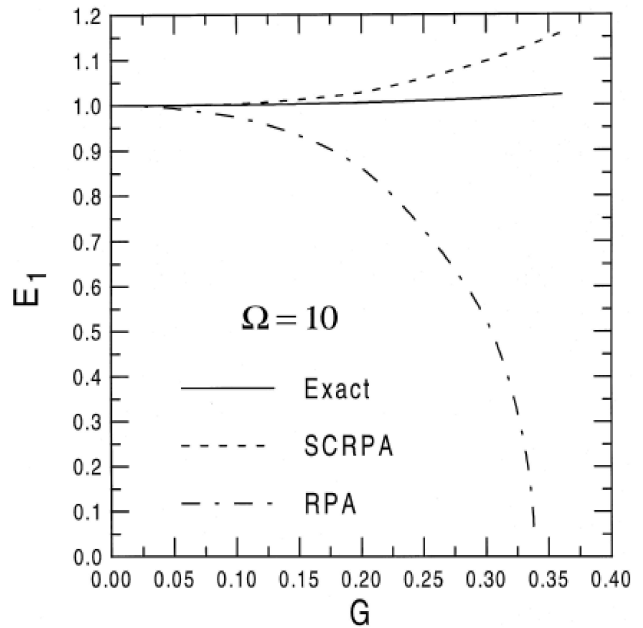}}
\vskip-2cm
\caption{First excited state in SCppRPA for N =10 as a function of the pairing coupling strength $G$. 
The strong improvement over standard ppRPA should be observed. The exact result is presented by the full line}
\label{E1} 
\end{figure}

\begin{figure}
\resizebox{0.50\textwidth}{!}{%
\includegraphics{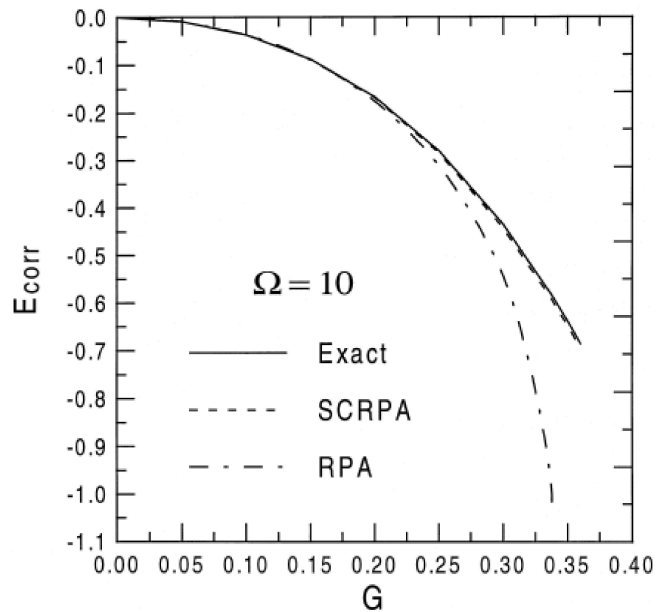}}
\vskip-2cm
\caption{Correlation energy for 10 particles from SCppRPA compared with the exact result as function of coupling}
\label{Ec} 
\end{figure}

In Figs. \ref{E1} and \ref{Ec}, we show results obtained with this factorised approximation for the first excitation energy with $N$ =10 particles and the correlation energy, respectively, as a function of the pairing coupling
strength $G$.
The exact results were obtained from the equations established by Richardson almost half a century ago \cite{Richardson}.

Commenting on the SCRPA results, we see that they 
are very much improved over standard ppRPA, \cite{Rin80}. On the other hand, we also see from the sum rule relation  given in \cite{Hir02}, Eq. (69), i.e. $\sum_{pp'}\langle N_pN_{p'}\rangle = \sum_{ph}\langle (2-N_h)N_p\rangle$ and Tables VII and XI in \cite{Hir02} that the  
Pauli principle, is still slightly violated, of the order of $4-5$ percent, what stems from the  fact that  the killing condition (\ref{killing}) is not exactly fulfilled.
Let us mention that SCRPA was solved in \cite{Hir02} and \cite{Duke-PLB}
among others for the case of 100 levels where it is even difficult to solve the 
problem with the Richardson equations.  A very instructive example is the $N=2$, i.e., the single Cooper pair case. Though already presented in \cite{Hir02}, let us discuss it here again. In standard RPA the excitation energy is given by

\[ E \propto \sqrt{1-G}\]

\noindent
whereas in SCRPA the result is

\[ E \propto \sqrt{1+G}~.\]

\noindent
The latter coincides, as already mentioned, with the exact result. The RPA result shows the usual BCS instability at $G=1$. With SCRPA the vertex renormalisation from the self consistency, i.e. screening, has effectively turned the sign of G around and with screening the effective interaction is now {\it repulsive} ! This stems from the fact that for $N=2$ the constraint from the Pauli principle is, as one easily realises, of maximum importance. It is testified again with this example that the Pauli principle is very well respected with SCRPA.

Let us also mention that the SCRPA scheme has been generalised to finite temperatures using an equivalent Green's function formalism in an application to the PFM in \cite{Sto03} with the same quality of results as at zero temperature. In particular it could be shown that also in the PFM, there opens a 
pseudo gap in the level density  approaching the critical temperature from above.

\subsection{Goldstone mode within a 3-level spin model and SCRPA}

The SCRPA scheme is a selfconsistent in-medium two body equation of the Schroedinger type. We think it is amenable for numerical solution for realistic systems. However, this needs major investments and this is not available at this moment. We, therefore, have chosen a simplified model to demonstrate numerically the fullfillment of the Goldstone mode. We have chosen the three level Lipkin model which also can be seen as a three sites spin model, corresponding to a SU(3) algebra \cite{Hol74}.
This model has been  used in order to test different many body approximations \cite{Hag00,Gra00}. We also have treated it already in \cite{Del05} with results of similar quality as in the two preceding models of sections VI.1 and VI.2. 
Here we want to dwell specifically on the zero mode. It is so far the only model where the appearance of the Goldstone (zero) mode has been demonstrated in a numerical application with SCRPA.

By labeling the levels 
with $0,~1,~2,$ we consider
the following Hamiltonian written in some 'original' basis 

\begin{equation}
H=\sum_{k =0}^2\epsilon_{k}S_{kk} -\frac{V}{2}\sum_{k =1}^2(S_{k 0}S_{k 0} + S_{0 k}S_{0 k})~,
\label{HL3}
\end{equation} 

\noindent
where $\epsilon_0=0;~~ \epsilon_1 =1;~~ \epsilon_2 = 1 + \Delta \epsilon~$ and

\be
S_{k k'} = \sum_{\mu =1}^Na^{\dag}_{k \mu}a_{k' \mu}
\label{Ksph}
\ee

\noindent
We suppose that the three levels have equal degeneracy $N$ and that in the non-interacting case the lowest level is full so that $N$ corresponds also to the particle number. The operators (\ref{Ksph}) satisfy simple commutation relations

\begin{equation}
[S_{k_1k_2},S_{k_3k_4}] = \delta_{k_2k_3}S_{k_1k_4} - \delta_{k_1k_4}S_{k_3k_2}~.
\label{Kcomm-rel}
\end{equation}

Standard HF-RPA shows a zero mode in the so-called deformed region where HF in the original basis  is unstable and when the two upper level become degenerate ($\Delta \epsilon =0$). This because the hamiltonian commutes with the 'angular momentum' operator $\hat L_0 = i(S_{21}-S_{12})$. According to \cite{Hol74}, the transformation matrix  to the deformed basis in (\ref{H}) can be written as follows

\begin{equation} 
R^{\dag}_{\alpha k}=
\left(\matrix{\cos \phi&\sin \phi&0 \cr -\sin \phi &\cos \phi &0 \cr 0&0&1}\right)~.
\label{trafo}
\end{equation}

\noindent
This means that we only have the single parameter $\phi$ to be varied to obtain the 'deformed' solution. First, let us write down the RPA operator in the deformed basis

\bea
Q^{\dag}_{\nu} &=& X^{\nu}_{10} A_{10} + X^{\nu}_{20} A_{20} + X^{\nu}_{21} A_{21}
\nn
 &-& Y^{\nu}_{10} A_{01} - Y^{\nu}_{20} A_{02} - Y^{\nu}_{21} A_{12}~,
\label{defQ}
\eea

\noindent
where the $A$ operators correspond to the $S$ ones of (\ref{Ksph}) but in the deformed basis.

\begin{figure}
\resizebox{0.50\textwidth}{!}{%
\includegraphics{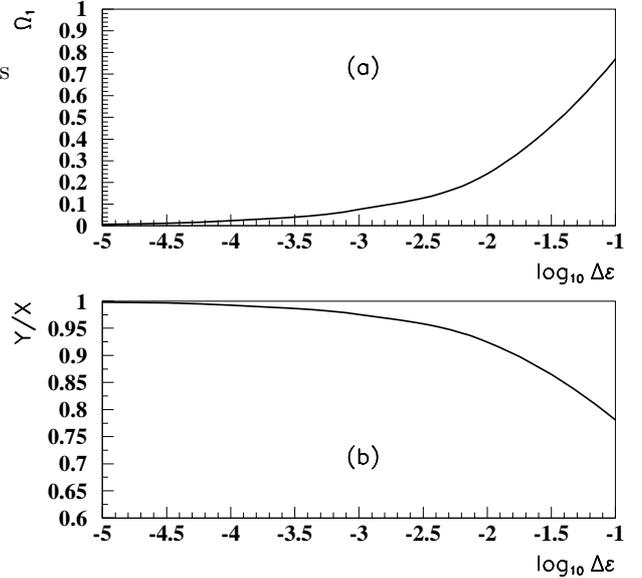}}
\caption{(a) First SCRPA excitation energy for $N=20$ and
$\epsilon_1=0,~\epsilon_2=1,~\epsilon_3=1+\Delta\epsilon $
versus $\log_{10}\Delta\epsilon$ (in arbitrary units).
(b) The ratio $Y/X$ versus $\log_{10}\Delta\epsilon$.}
\label{fig1}
\end{figure}

We also write the Hamiltonian in the deformed basis and then construct the SCRPA equations from the double commutator relations 
(\ref{A-matrix}, \ref{B-matrix}). While solving the SCRPA equations we have to minimise the ground state energy with respect to $\phi$ in order to fullfill the generalised mean field equation (\ref{rpa-mfbis}). The latter can be calculated as a function of $X, Y, \phi$ in expressing its expectation values from the inversion of (\ref{defQ}) and then using (\ref{killing}). Our procedure works with collective generators what seems natural in this model, since we can suppose that the non-collective states decouple to a large extent from the rest of states. Working with the individual quantum states (m-scheme) would considerably complicate the solution of the model with the SCRPA scheme. Employing the collective operators has, however, the disadvantage that expectation values of products of diagonal operators as $\langle A_{00} \rangle, \langle A_{00}A_{22} \rangle$, etc. cannot directly be expressed with the $X, Y$ amplitudes. We, therefore, in \cite{Del05}
 found expressions via the unitary operator method what yields

\bea
n_{\alpha}&\approx&\bigg [ y_{\alpha \alpha} + y_{11}y_{22}/N \bigg ]
\nn&\times&
\bigg [ 1 + 2(y_{11}+y_{22})/N + 3y_{11}y_{22}/N^2 \bigg ]^{-1}
\nn&&
\alpha = 1, 2~,
\label{occ-numbs}
\eea

\noindent
where $y_{\alpha \beta}=\sum_{\nu}Y^{\nu}_{\alpha 0}Y^{\nu}_{\beta 0}$.
Here, evidently $ n_0 = N - n_1 - n_2$ and  $n_{\alpha} = \la A_{\alpha \alpha} \ra$. 
For the quadratic terms we obtain \cite{Del05}
\bea
&&\la A_{00}A_{\alpha\alpha}\ra\approx\frac{N-1}{N}\la A_{\alpha 0}A_{0\alpha}\ra
\nn&-&
\frac{1}{N^2(N-1)}(\la A_{\alpha 0}^2A_{0\alpha}^2\ra+\la A_{10}A_{20}A_{02}A_{01}\ra)
\nn
&&\la A_{11}A_{22}\ra\approx\frac{1}{N(N-1)}\la A_{10}A_{20}A_{02}A_{01}\ra~.
\nn
\label{expans0}
\eea

Evaluating the four body correlation functions with the inversion (\ref{inversion}) and the killing condition (\ref{killing}), one obtains a set of linear equations for the two body correlation functions with diagonal $A$ opertors which can be solved. With this, we  get the solution of the SCRPA equations in the deformed regime.
In Fig. \ref{fig1} (a) we show the energy of the first excited state as a function of
 $\Delta \epsilon = \epsilon_2 - \epsilon_1$ (in arbitrary units). 
As already mentioned, in the limit $\Delta \epsilon \rightarrow 0$ a zero mode should appear. We see that there is a rapid decrease of the first excitation energy $\Omega_1$ to zero. 
Of course the whole system of SCRPA equations is very sensitive to numerical accuracy. 
For instance the minimum of the ground state energy at $\phi = \phi_0$ is not easy to determine with high accuracy
which we estimate to be of order $10^{-3}$. With this, for $\Delta \epsilon = 10^{-5}$, 
we obtain an excitation energy of order $10^{-2}$. It also is interesting to follow the values of the $X, Y$ amplitudes. 
In Fig. \ref{fig1} (b) we show the evolution of their ratio $Y/X$ as it approaches one. 
At exactly zero excitation energy the amplitudes would diverge. 
Here, they are still of reasonable value, i.e. $X\sim Y\sim$ 20.
It is worth mentioning that even a very tiny inaccuracy in $\phi_0$ destabilises the zero mode showing that it is absolutely neccessary to work in the basis which fullfills eq. (\ref{rpa-mfbis}). The scenario stays more or less the same, if instead of SCRPA the simpler r-RPA is applied (see Section III).
We think that this is a very instructive example which clearly demonstrates that SCRPA in the form presented here
with all components included conserves all appreciated properties of standard HF-RPA. To the best of our knowledge, we do not know of any other method which with more or less equal performances obtains the zero mode in a similarly easy way. Even for this simple model the realisation of the zero mode with the $\Phi$ derivable functional \cite{Kad61} would be very combersome.

\section{A short outline of the equivalent Green's function description of the 
equation of motion method}

In condensed matter physics dealing with homogeneous infinite systems, one usually does not formulate the problems in the form of an eigenvalue equation. One rather employs propagators or many body Green's functions. Of course, it is clear that every eigenvalue problem has a corresponding formulation with Green's functions but it may be useful to give some more  details on the ingredients of the present formalism. However, the Green's function equivalent to the eigenvalue equation of SCRPA 
(\ref{rpa-eq}) is, in a way,  somewhat particular. As one may immediately realise, it cannot come from the familiar many time Green's function approach where, e.g., the two body propagator (and also its integral kernel) depends on four times once one goes beyond the standard HF-RPA scheme. This stems from the fact that in an eigenvalue problem only one energy (the eigenvalue) is involved and then the corresponding integral equation for the Green's function also can involve only one energy, even in the integral kernel.  Though the formalism has been described in earlier publications, see, for instance, refs. \cite{Sto03} and \cite{Duk98},
we feel that it may be helpful for the reader to give a short outline of the procedure. 
To this purpose, we write down the corresponding integral equation form of (\ref{short-hand-rpa}), that is the Bethe-Salpeter equation

\begin{eqnarray}
&(&\hbar \omega - E_{k_1} +E_{k_2})\tilde{\mathcal G}^{\omega}_{k_{1}k_2k_{3}k_4} \nonumber\\
&=& {\mathcal N}_{0, k_1k_2}[
\delta_{k_1k_3}\delta_{k_2k_4} +\sum_{k_{3'}k_{4'}}   {\mathcal S}_{k_1k_{2}k_{3'}k_{4'}}]\tilde{\mathcal G}^{\omega}_{k_{3'}k_{4'}k_{3}k_4}
\label{BSE}
\end{eqnarray}

Inserting the spectral representation for the Green's function

\begin{equation}
\tilde{\mathcal G}^{\omega} = \sum_{\mu}\frac{\chi^{\mu}N_{\mu}\chi^{\mu^*}}{\hbar \omega - \Omega_{\mu} + i\eta N_{\mu}}
\label{spectralGF}
\end{equation}
where the sum goes over positive and negative values of $\mu$ and $N_{\mu} = - N_{-\mu}, \Omega_{\mu} = -\Omega_{-\mu}$,
and taking the limit $\hbar \omega \rightarrow \Omega_{\nu}$, we obtain in comparing the singularities on left and right hand sides, the eigenvalue equation (\ref{short-hand-rpa}). 

In order to see how this scheme with the equation of motion technique can go on and lead to an $\omega$-dependent term in the integral kernel of the Bethe-Salpeter equation, we extend the opertor (\ref{rpa-op}) to include a two body term as a first extension, eventually higher order terms

\begin{equation}
\tilde Q_{\nu}^{\dag} = \sum[\tilde \chi_{\alpha \beta}a^{\dag}_{\alpha}a_{\beta} + \tilde {\mathcal X}_{\alpha \beta \gamma \delta}a^{\dag}_{\alpha}a^{\dag}_{\beta}a_{\delta} a_{\gamma} + ...]
\end{equation}

This leads to an extended eigenvalue problem involving also the two body amplitudes ${\mathcal X}$. Eliminating the latter from the coupled equations of one body and two body amplitudes, one obtains an effective equation for the $\chi$ amplitudes with an effective, energy dependent potential containing implicitly the two body amplitudes. This effective potential can be qualified to corresponds to the $\omega$ dependent part of a two body self energy. This procedure can formally be pushed up to the N-body amplitudes leading thus to an exact two body  Dyson equation form, in analogy to what is known from the single particle Green's function.

Let us shortly show how the same scheme can be obtained beginning directly with the Green's function. We start with the following chronological propagator


\begin{equation}
{\mathcal G}_{12}^{t-t'} = -i \langle 0|\mbox{T}A_1(t)A_{2}^+(t')|0\rangle,
\label{defG}
\end{equation}

\noindent
with $A(t) = e^{iHt}A(0)e^{-iHt}$, T the time ordering operator and


\[A_1 = a^{\dag}_{k_{1'}}a_{k_1}~,~~~~ A_{2}^{\dag}= a^{\dag}_{k_2}a_{k_{2'}} \]

\noindent
where $a^+, a$ are fermion creation and destruction operators, respectively and the Green's function in (\ref{defG}) is thus a density-density correlation function. It is always understood that the indices $k_i$ comprise, as before, momentum and spin and, eventually more quantum numbers, such as isospin, etc. We remark that in this definition of the Green's function we put pairs of fermion operators on equal times so that the Green's function depends only on one time difference at equlibrium. 
The $\tilde {\mathcal G}$ Green's function (\ref{spectralGF}) is related to ${\mathcal G}$ of (\ref{defG}) in replacing in the latter the $A_1$ by $\tilde A_1 = a^{\dag}_{k_{1'}}a_{k_1}/\sqrt{M_{k_{1'}k_1}}$, etc.\\
We now claim that for the two time Green's function (\ref{defG}), one can write down in a well defined way a formally exact integral equation with an integral kernel which also depends only on one time difference (or in energy space on one energy $\hbar \omega$). We, thus, write 

\begin{equation}
{\mathcal G}^{\omega} = {\mathcal G}^{\omega}_0 + {\mathcal G}^{\omega}_0\Sigma^{\omega}
{\mathcal G}^{\omega}~,
\label{BSGF}
\end{equation}

\noindent
where it is understood that this is a matrix equation with matrix multiplication of the various products. The lowest order Green's function ${\mathcal G}_0$ is thereby given for, e.g., a translationally invariant system as

\begin{equation}
{\mathcal G}_{0,12}^{\omega} =\frac{n_{k'_1} - n_{k_1}}{\hbar \omega -E_{k_1} + E_{k'_1}}\delta_{k_1k_2}\delta_{k'_1k'_2}
\label{G_0}
\end{equation}

\noindent
where $n_k = \langle 0|a^{\dag}_ka_k|0\rangle$ are the single particle occupation numbers and $E_k = k^2/(2m) + \sum_{k'}\bar v_{kk'kk'}n_{k'}$ are the mean field energies.

In principle, Eq. (\ref{BSGF}) may thus serve as a definition of the kernel $\Sigma^{\omega}$. It turns out that $\Sigma^{\omega}$ is a well defined object for which expressions in terms of usual correlation functions and Green's functions can be given, see, e.g., \cite{Duk98}. This kernel can be considered as some kind of higher order self energy, here the self-energy of density fluctuations. As the well known self-energy of the single particle Green's function, it splits into an instantaneous, energy independent part $\Sigma^0$ and an explicitly energy dependent part $\Sigma^r(\omega)$. It can be shown that $\Sigma^0$ is equivalent to the matrix ${\mathcal S}$ in (\ref{short-hand-rpa}) as this is explained in \cite{Duk98}. Therefore (\ref{short-hand-rpa}) and (\ref{BSGF})) are equivalent once $\Sigma^{\omega}$ is replaced by its static part $\Sigma^0$. Mathematically, this can be seen quite straightforwardly in applying the equation of motion to the propagator (\ref{defG}): $i\frac{\partial}{\partial t}{\mathcal G}_{12} = \delta(t-t')\langle 0|[A_1,A^+_2]|0\rangle -i\langle 0|\mbox{T}[A_1,H]_tA^+_2(t')|0\rangle$. Applying now the equation of motion a second time to the time $t'$ figuring in the correlation function which appears on the r.h.s. of this equation, one realises that the part which acts on the chronological operator $\mbox{T}$ leads to the double commutator also involved in ${\mathcal S}$ of Eq.(\ref{short-hand-rpa}) and, consequently, in the instantaneous part of the self energy $\Sigma^0$. The application of the time-derivative on $t'$ contained in $A^+_2(t')$ will lead to the energy dependent part of the self-energy in (\ref{BSGF}). This brief outline should only serve to give the reader a quick feeling how such a somewhat unusual integral equation like (\ref{BSGF}) with an integral kernel depending only on one energy can be obtained. For a more detailed outline, we refer the reader to \cite{Duk98}. \\

Concerning the practical solution of (\ref{BSE}), it can be seen
 from (\ref{short-hand-rpa}), that the static part only contains up to two body correlation functions which can be calculated from (\ref{BSE}) and, thus, a selfconsistent cycle is established. As just explained, the dynamic, explicitly energy dependent part contains the coupling to higher configurations involving four body propagators. Their inclusion leads in some approximation to what is known in the equation of motion method as the second RPA equations.\\

\begin{figure}
\resizebox{0.35\textwidth}{!}{%
\includegraphics{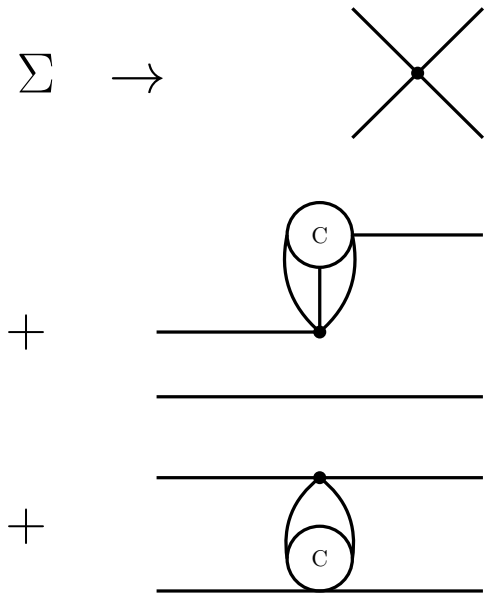}}
\caption{Schematic representation of the two body mean field. It consists of the direct interaction (full dot), the single particle renormalisation due to  two body correlations ($C$), and screening terms where a two body correlation is echanged between the two lines. It should be mentioned that the two body renormalisations occur instantaneously as indicated when the time runs from left to right. Also symmetric terms should be added interchanging upper and lower lines. The tad-pole graphs are supposed to be included in the single particle lines.}
\label{fig2}
\end{figure}

It may be worth mentioning that a perturbative analysis of $\Sigma$ in (\ref{BSGF})) shows that the terms are not equivalent to Feynman diagrams. Nevertheless, one can present the various terms in $\Sigma^0$ (or equivalently in ${\mathcal S}$ of eq (\ref{short-hand-rpa})) by the graphs shown in Fig. \ref{fig2}. If in this figure the two body correlation functions are replaced by the interaction, the standard second order perturbation graphs emerge with, however, the particularity that they occur instantaneously, that is they do not propagate. Even, if the correlation functions in Fig. \ref{fig2} are replaced by their full expression, the graphs stay, as indicated in the figure, instantaneous. This feature results from the minimisation of the energy weighted sum rule as explained in section 2.\\

Similar type of equations with integral-kernals depending only on one frequency are obtained from the hypernetted chain equations, see \cite{Saa08}.

\section{ Discussion and Conclusions}

In this work we summarized some results obtained with SCRPA. With respect to earlier publications, we elaborated on a particular aspect of the Self-Consistent RPA (SCRPA) approach for cases with a continuously broken symmetry which concerns the appearance of a Goldstone (zero) mode and the fullfillment of the energy weighted sum-rule. This SCRPA sums up correlations beyond standard RPA and it is, a priory, not evident that the desirable properties of standard RPA such as being a conserving approximation implying fullfillment of energy weighted sum rule and appearance of Goldstone (zero) modes are maintained. 
In fact it is known that maintaining these properties in beyond standard RPA approaches is particularly difficult and a strongly debated subject in the literature, see e.g. \cite{Yan15} for a recent publication.
We demonstrate in this work, theoretically and, for a model case, numerically that, indeed, SCRPA maintains all those desirable properties for practically all situations of spontaneously broken symmetries with the symmetry operator being of the one body type. An exception occurs for pairing in finite systems because in the symmetry operator (the number operator) appears a hermitian piece which cannot be incorporated in the SCRPA formalism. However, in homogeneous infinite systems, this again causes no essential problem. Another important quality of standard RPA concerns electromagnetic gauge invariance. This was not the subject of this work, however, let us shortly discuss how SCRPA is coping with this issue. For example, in an instructive paper by Feldman and Fulton \cite{FF}, it is demonstrated in transparent terms how standard RPA fulfils gauge invariance. This property is also maintained with SCRPA. The extra terms containing the two body correlation functions in (\ref{short-hand-rpa}) cancel in the limit where the two open legs are put on the same spot in position space. Actually, gauge invarince of standard RPA {\it as well as} SCRPA can easily be verified from (\ref{A-matrix},\ref{B-matrix}). If in these equations the operator $\delta Q_{\alpha \beta}$ is transformed into $r$-space and the diagonal element is taken, as demanded to show gauge invariance (see \cite{FF}, Eq. (3.69)), we immediately realise that this diagonal operator commutes with the remainder (also written in $r$ space), once the Hamiltonian $H$ is replaced by its interaction part $V$, that is, the Coulomb interaction. Therefore,  gauge invariance is fullfilled. This argument is valid discarding spin but, as shown in \cite{FF}, this does not invalidate the general proof. These considerations also entail that the so-called 'velocity-length' equivalence in the dipole transition is preserved \cite{FF}.\\
 We outlined the SCRPA equations first as an eigenvalue problem but also showed how the equivalent Green's function approach can be formulated. The SCRPA equations are selfconsistent two body equations of the Schroedinger type obtained variationally from the minimisation of the energy weighted sum-rule which should be amenable to numerical treatment for realistic problems. In the model cases treated here and in other earlier publications, the results are generally of very good quality improving substantially over standard RPA for instance in the vicinity of a phase transition point.

{\bf Acknowledgments}

We are greatful to Markus Holzmann for discussions and his interest in this work. We thank  J. Dukelsky and M. Jemai for interesting discussions
and past collaborations and A. Dumitrescu for technical support. Helpful remarks by Kurt Schoenhammer on the manuscript have been appreciated.
This work was supported by the agreement between IN2P3 and IFIN-HH
and the grants of the Romanian Ministry of Education and Research,
PN-II-ID-PCE-2011-3-0092, PN-09370102.

The Authors equally contributed to the paper.




\begin{thebibliography}{99}
\bibitem{Nig99} M.P. Nightingale, C.J. Umrigar, (Eds.) 
{\it Quantum Monte Carlo Methods in Physics and Chemistry}, (Springer, Berlin, 1999).
\bibitem{Hol11} M. Holzmann, B. Bernu, C. Pierleoni, J. Mc Minis, D. M. Ceperly, V. Olevano, L. Delle Site, Phys. Rev. Lett. {\bf 107}, 110402 (2011)
\bibitem{Pes99} I. Peschel, X. Wang, M. Kaulke, and K. Hallberg (Eds.), 
{\it Density-Matrix Renormalization, A New Numerical Method in Physics}, (Springer, Berlin, 1999).
\bibitem{Sch11} U. Schollwoeck, Ann. Physics {\bf 326}, 96 (2011) {\it and} Rev. Mod. Phys. {\bf 77}, 259 (2005)
\bibitem{Row68a} D.J. Rowe, Rev. Mod. Phys. {\bf 40}, 153 (1968).
\bibitem{rpae} We want to clarify a semantic point: in our nomenclature RPA stands for RPA including the ladder (Coulomb) diagrams. In the condensed matter literature, this is sometimes called RPAC. However, in other fields, like nuclear physics or physical chemistry, RPA stands  for linearised time-dependent HF. Since our extension of RPA(C), the SCRPA, has in the past been used in condensed matter and nuclear theory including ladders, we will keep the names RPA and SCRPA including the Coulomb diagrams.
\bibitem{Kad61} G. Baym and L.P. Kadanoff, Phys. Rev. {\bf 124}, 287 (1961);
G. Baym, Phys. Rev. {\bf 127}, 1391 (1962);
G. Baym, Phys. Lett. {\bf 1}, 242 (1962).
\bibitem{Del05} D.S. Delion, P. Schuck, and J. Dukelsky
Phys. Rev. C {\bf 72}, 064305 (2005).
\bibitem{Duk90} J. Dukelsky and P. Schuck,
Nucl. Phys. {\bf A512}, 466 (1990).
\bibitem{Hir02} J.G. Hirsch, A. Mariano, J. Dukelsky, and P. Schuck,
Ann. Phys. (NY) {\bf 296}, 187 (2002).
\bibitem{Cat96} F. Catara, G. Piccitto, M. Sambataro, and N. Van Giai,
Phys. Rev. {\bf B54}, 17536 (1996);\\
F. Catara, M. Grasso, G. Piccitto, and M. Sambataro,
Phys. Rev. {\bf B58}, 16070 (1998).
\bibitem{Sto03} A. Storozhenko, P. Schuck, J. Dukelsky, G. R\"opke,
and A. Vdovin, Ann. Phys. (NY) {\bf 307}, 308 (2003).
\bibitem{Rin80} P. Ring and P. Schuck,
{\it The Nuclear Many-Body Problem} (Springer Verlag, New York, 1980).
\bibitem{Duk98} J. Dukelsky, G. R\"opke, and P.Schuck,
Nucl. Phys. {\bf A628}, 17 (1998).
\bibitem{Jem13} M. Jemai, D.S. Delion, and P. Schuck, 
Phys. Rev. C {\bf 88}, 044004 (2013).
\bibitem{Bla86}J.P.Blaizot and G. Ripka,
{\it Quantum Theory of Finite Systems}
(MIT Press, Cambridge, 1986).
\bibitem{Gol61} J. Goldstone, Nuovo Cimento {\bf 19}, 154 (1961).
\bibitem{Koh61} W. Kohn, Phys. Rev. {\bf 123}, 1242 (1961).
\bibitem{Rei01} J. Reidl, G. Bene, R. Graham, P. Sz\'epfalusy, Phys. Rev. A {\bf 63}, 043605 (2001).
\bibitem{And58} P.W. Anderson. Phys. B {\bf 112}, 1900 (1958);
N.N. Bogoliubov. Sov. Phys. JETP, {\bf 34}, 698, (1958).
\bibitem{Com06} R. Combescot, M. Yu. Kagan, and S. Stringari,
Pys. Rev. A {\bf 74}, 042717 (2006).
\bibitem{Urb14} No\"el Martin and Michael Urban,
Phys. Rev C {\bf 90}, 065805 (2014).
\bibitem{Bes66} D.R. Bes and R.A. Broglia,
Nucl. Phys. {\bf 80}, 289 (1966).
\bibitem{Kha04} E. Khan, N. Sandulescu, Nguyen Van Giai, and M. Grasso,
Phys. Rev. C {\bf 69} 014314 (2004).
\bibitem{Tohy15} M. Tohyama and P. Schuck, Phys. Rev. C {\bf 91}, 034316 (2015).
\bibitem{Mohsen} M. Jemai, P. Schuck, J. Dukelsky, R. Bennaceur, Phys. Rev. B {\bf 71},085115, 2005.
\bibitem{Luttinger} M. Urban, P. Schuck, Phys. Rev. A {\bf 90}, 023632, 2014
\bibitem{Schaefer} S. Schaefer, P. Schuck, Phys. Rev. B {\bf 59}, 1712, 1999.
\bibitem{Duke-PLB} J. Dukelsky, P. Schuck, Phys. Lett B {\bf 164}, 164, 1999.
\bibitem{Richardson} R. W. Richardson, Phys. Rev. {\bf 141}, 949, 1966.
\bibitem{Hol74} G. Holzwarth and T. Yukawa,
Nucl. Phys. {\bf A219}, 125 (1974).
\bibitem{Hag00} K. Hagino and G.F. Bertsch,
Phys. Rev. {\bf C61}, 024307 (2000).
\bibitem{Gra00} M. Grasso and F. Catara,
Phys. Rev. {\bf C63}, 014317 (2000).
\bibitem{Saa08} M. Saarela, Lecture notes at fall 2008, University of Oulu; Material for reading: A. Fabrocini, S. Fantoni, and E. Krotscheck (Eds): {\it Introduction to Modern Methods of Quantum Many-Body Theories and Their Applications}, Series on Advances in Quantum Many Body Theory-Vol. 7, World Scientific, London, (2002).
\bibitem{Yan15} Yan He and Hao Guo, arXiv:1411.5100.
\bibitem{FF}
G. Feldman, T. Fulton, Ann. Phys. {\bf 152}, 376 (1984).


\end{thebibliography}
\end{document}